# Method Article

**ABSTRACT**

*We present here the straightforward implementation of pump-probe methods into existing scanning tunneling microscopy (STM) systems. Our method uses the waveform-sequencing abilities of a standard arbitrary waveform generator (AWG) and a simple mechanical relay switch that either connects the regular STM control electronics or the AWG to the STM system. Our upgrade further enables pulsed-ESR excitation for advanced STM based spin-resonance experiments. We demonstrate the technical implementation, signal detection using a lock-in amplifier, and cross-correlation measurements of DC/DC and DC/RF pulses highlighting our ~5 ns time resolution, here limited by the speed of the available electronics. Our setup is highly versatile and can be extended to suit other needs of STM based investigations such as required in diverse mapping schemes or the coherent manipulation of qubits.*

- *Plug and Play Pump-Probe Capability*
- *One setup for Pump-Probe Spectroscopy and Pulsed-ESR*
- *Waveform Memory Saving and Versatile due to Waveform-Sequencing of Arbitrary Waveforms*

**SPECIFICATIONS TABLE**

| Subject Area | Select one of the following subject areas:<br>• Physics and Astronomy |
|---|---|
| More specific subject area: | *Surface physics, time-resolved measurements* |
| Method name: | *Waveform-sequencing for scanning tunneling microscope based pump-probe spectroscopy and pulsed-ESR* |
| Name and reference of original method | *Measurement of the Fast Electron Spin Relaxation Times with Atomic Resolution, Loth et al., Science* **329**, *1628 (2010)* |

**Method details**

Broader Context

The small signals and the concomitantly required amplification through transimpedance amplifiers limit the bandwidth of scanning tunneling microscopes (STM) to the kilo Hertz range. Although specialized circuit designs may achieve MHz bandwidths [1], they operate at cryogenic temperatures which renders them difficult to implement, thus limiting their versatility.

The introduction of pump-probe techniques for STM systems by Nunes and Freeman [2,3] more than two decades ago lifted amplifier related constraints of the bandwidth and promised unprecedented access to the temporal dynamics of atomic scale systems. However, these earlier approaches and more recent ones using THz pulses [4,5] require either dedicated sample and tip designs or a sophisticated external setup for creating pulses and their coupling into the STM junction.

The elegant all-electronic pump-probe spectroscopy method, introduced by Loth and co-workers [6], enabled investigation of spin relaxation times in the nanoseconds range with atomic resolution. It has since served to measure the dynamics of long-lived antiferromagnetic structures [7,8] and individual atoms [6,9–13]. In these pump-probe setups the existing bias voltage line of an STM, a dedicated pulse pattern generator, and a mechanical chopper to modulate the probe-pulses for lock-in detection [6,14] are used. It is surprising that even though this solution appears quite straightforward, it has not seen a more widespread implementation by the community.

We therefore introduce here an even simpler setup for obtaining pump-probe capabilities in the nanoseconds range using only a standard arbitrary waveform generator (Keysight 33600A) and its waveform-sequencing function. Our approach is truly *plug-and-play*, affordable, and should alleviate any remaining hesitations encountered by other researchers in the planning of such an addition to their existing STM apparatus.

Method details

The integration of pump-probe capabilities is done within minutes by simply adding an in-line relay to the bias voltage path [15]. The schematics in Figure 1 shows the hookup of the AWG (red) to the STM control electronics and the optional RF generator (dashed yellow). In contrast to previous pump-probe setups where an RF relay served as a mechanical chopper for lock-in detection, we use the relay as a transfer switch to either connect the DC bias from the STM control electronics (for tunneling spectroscopy and regular topographic imaging) or the pulses from the AWG to the STM apparatus [15]. For the present purpose, we use a regular PCB mount relay (Tianbo) that we control with a 5V USB power supply and to which we solder BNC connectors. Different relays of any sophistication may equally work but are not needed in our case, since neither edge times are affected nor is any ringing introduced by the relay.

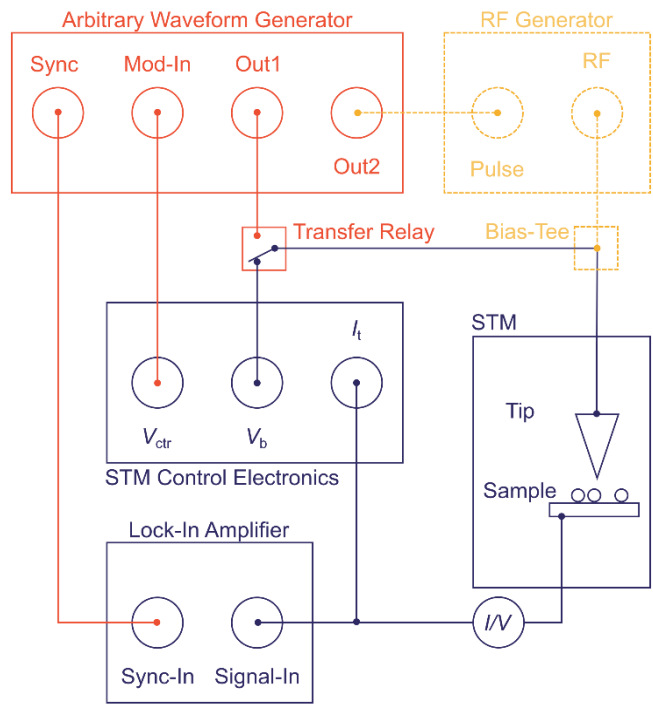

*Figure 1: Blue drawings indicate existing STM system, control electronics, and lock-in amplifier. The red sketch shows how to implement the AWG into that system using a transfer relay. The dashed yellow drawing is the optional integration of a pulsed RF generator.*

Our method for pump-probe spectroscopy and pulsed-ESR uses the waveform-sequencing ability of the AWG. Waveform-sequencing is essentially the consecutive and seamless execution of a list of predefined arbitrary waveforms (arbs), often compared to playing songs from a playlist in a music player.

Figure 2a illustrates our method. We first create three arbs for the pump-pulse (arbP: yellow) and probe-pulses (arbA: green, and arbB: orange) using MATLAB. These arbs have equal sample length of $l = t_{arb} \times SR$, where *SR* is the sample-rate of the AWG. The length of the arbs should be designed with the dynamics of the studied system in mind. This means that the total duration of the arb, i.e., $t_{arbA}$, should be several times the expected relaxation times to ensure excitation of the system from a known equilibrium state. The two distinct probe-pulse arbs (arbA and arbB) are applied cyclically in Ch1 at about $f_{cycle}$ = 887 Hz to modulate the tunneling current for lock-in detection. Our waveform playlist is composed of a dedicated arb sequence of probe-pulses consisting of *N* segments of arbA (cycle A) followed by *N* segments of arbB (cycle B), where $N = \lfloor 1/2 f_{cycle} t_{arbA} \rfloor$. To provide a synchronization signal for the lock-in amplifier, we set the sync output to high (Hi) during cycle A, and during cycle B we set it to low (Lo). The effective $f_{cycle}$ will depend on the rounding down in the determination of *N*.

The execution of the pump-arb is simpler as we play it perpetually on Ch2, as also indicated in Figure 2a. The waveform-sequence of Ch1 and the pump-arbs on Ch2 are added internally by the AWG and finally applied to output 1. As an example, we show in Figure 2b a measured time-trace of this combined output 1 signal, recorded with an oscilloscope at the threshold between cycle A and B. The transition from one cycle to the other is as seamless as the transitions between any arbs. The three arbs used here have a duration of 50 ns. ArbP contains a 5 ns wide pulse of 150 mV amplitude and is generated by Ch2. The waveform-sequence composed of arbA and arbB is generated on Ch1 and contains a 10 ns wide pulse of 50 mV amplitude for arbA, while arbB =- arbA.

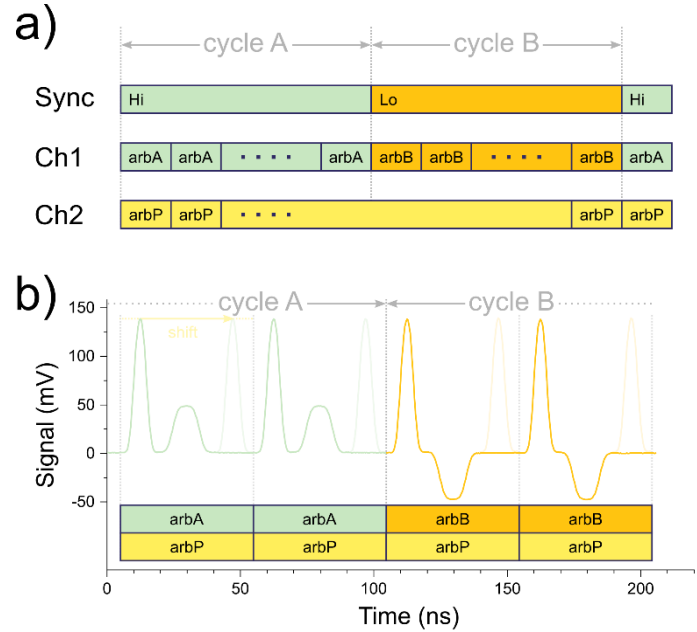

Figure 2: Waveform-sequencing concept. a) The waveform-sequence consist of two repeated cycles of distinct probe-arbs (arbA and arbB) and is played on Ch1. The pump-arb (arbP) is generated in Ch2. Both channels are added internally by the AWG and applied to output 1. A sync-signal marks the temporal position within the waveform-sequence for lock-in detection. b) Measured time-trace of output 1 at the end of cycle A, demonstrating the seamless transition between cycles. The shift of the probe-pulse is used in time-resolved measurements and serves to vary the time-delay between pump and probe-pulse.

We now discuss how to use this setup for time-resolved measurements. As we see from Figure 2b, the pulses constitute only a short part of their respective arb. Since our waveform-sequencing guarantees the perfect synchronization of Ch1 and Ch2, arbP will always be in sync with arbA and arbB. This leaves the pulse-position within the arb, i.e., the phase, as the handle through which to control the time-delay $\Delta t$ between pump and probe-pulses.

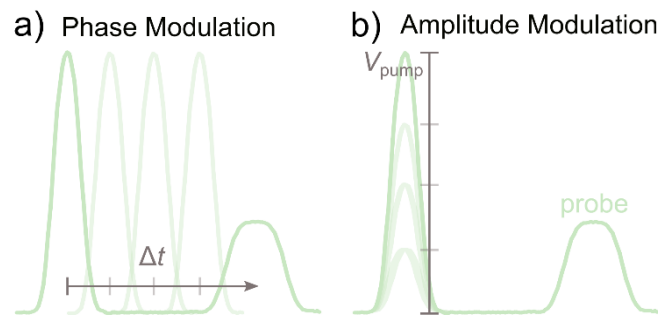

Figure 3: Modulation schemes acting on the pump-arb (arbP), played on Ch2. (a) The phase modulation (PM) adjusts the relative time $\Delta t$ between the pump and the probe-pulse. (b) The amplitude modulation (AM) controls the pump-pulse amplitude at fixed $\Delta t$. Both modes are controlled via the external $V_{ctr}$ voltage.

We conveniently vary the position of the pump-pulse within arbP via the phase modulation function (PM, see Figure 3a). To that end, we use the Mod-In input of the AWG to steer the phase of Ch2 by using an external DC voltage ($V_{ctr}$) from the STM control electronics (see schematics Figure 1). A dedicated DAQ card may serve that purpose as well. We adjust the phase of arbP by +/-180 degrees for $V_{ctr}$ = +/-5 V. This is equivalent to the circular shift of the pulse position within arbP when $V_{ctr}$ is swept from -5 to +5 V and corresponds to a variation of the relative time $\Delta t$ between pump and probe-pulse in the fully available range $t_{arbA}$. Note that the phase of Ch2 may also be adjusted via the computer control of the AWG, which might be useful for certain applications or when no control voltage is available.

Similarly, for a chosen time delay between pump and probe-pulse, we again use the amplitude modulation (AM, see schematics in Figure 3b) of the AWG to control the amplitude of the pump-pulses via $V_{ctr}$. A voltage of +5 V corresponds to the maximal amplitude and -5 V to the maximal negative amplitude. The AM mode is useful in the search for threshold voltages at which dynamical processes are excited, as for instance shown in previous work [6,9].

Method validation

To reveal the time-resolution of our method through the bottle neck of a bandwidth limited system, for instance a transimpedance amplifier, we exploit a nonlinear tunneling junction, which is frequently used as a base for time-resolved measurements [2]. We simulate here a tunneling junction by a Schottky diode (Pasternack, PE8014). The nonlinearity creates an additional current when pump and probe-pulses coincide due to $I(V_{pump} + V_{probe}) \neq I(V_{pump}) + I(V_{probe})$ [2]. Figure 4 shows the resulting diode signal when we overlap pump and probe-pulses generated by the earlier described waveform-sequencing method and the phase-modulation. At first, we show the overlap of DC pump and DC probe-pulses on the nonlinearity in panel a. We vary the duration of the pump-pulse from 5 to 15 ns at a fixed probe-pulse width of 10 ns to reveal the correlation effect of the nonlinearity.

We further use the same method to superimpose RF pump-pulses and DC probe-pulses, as seen in the resulting correlation trace in Figure 4b. Instead of internally combining Ch1 and Ch2 on output 1, we now feed Ch2 directly to the pulse-trigger input of the RF generator (Keysight N5183B). As indicated in Figure 1, the cabling remains conveniently untouched for this operation mode. The output of Ch1 is now solely playing the waveform-sequence of probe-pulses (arbA, arbB). The launched RF pulses travel via the resistive bias-tee (Tektronix, PSPL5370) to the tunneling junction where they overlap with the DC pulses. For this demonstration, we set the generator to a frequency of 10 GHz at a power level of 5 dBm. The probe-pulse width is held constant in Figure 4b while we vary the width of the triggered RF pulses from 5 ns to 20 ns. As can be seen in the correlation data, the RF pulses have the same width as their trigger pulses, thus presenting all requirements for pulsed-ESR methods that require the controlled application of RF pulses of precise timing and duration.

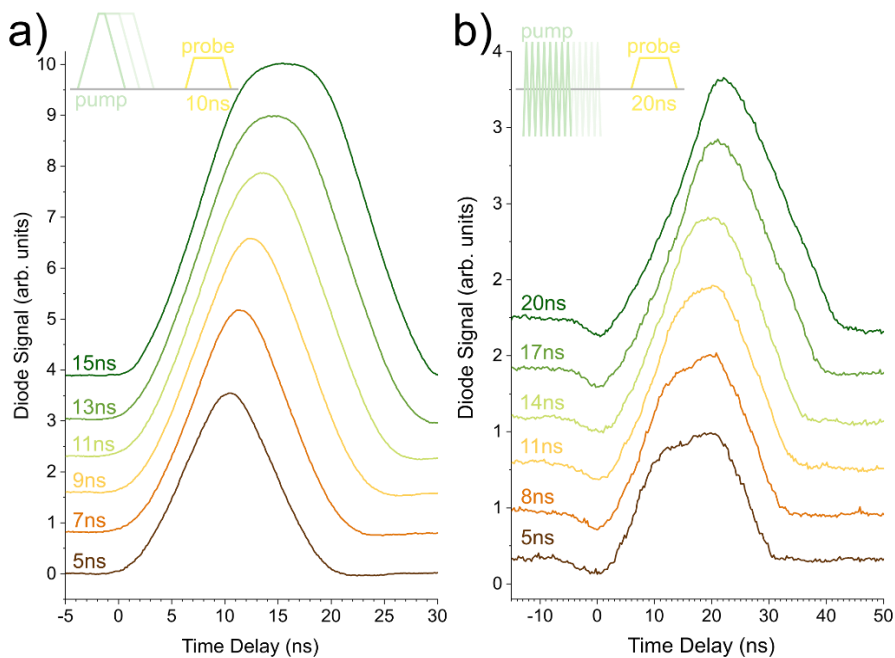

Figure 4: Cross correlation using a nonlinearity of (a) DC pump with DC probe-pulse (10 ns) and (b) RF pump with DC probe-pulse (20 ns). The pump-pulse width is indicated next to the respective correlation traces. The cross-correlation trace in panel (b) shows a subtle dip due to slight ringing introduced by the bias-tee.

The shape of the correlation curves in Figure 4 reveals important information about the effective time-resolution and offers valuable diagnostical insight into the system. For instance, the roundness of the correlation data is determined by the edge-times of the pump and probe-pulses, which is here about 3 ns and limited by the speed of our AWG. Further, the correlation data is useful in the detection of ringing due to the presence of impedance mismatches [16], which lead to the subtle dip in the Figure 4b and was caused by the bias-tee.

The time-delay of the maxima in the correlation data, such as in Figure 4, is extremely sensitive to variations in the cable lengths. To estimate the minimal length to which we are sensitive, we add cable segments of different lengths to the Ch2-Pulse path (Figure 1). The length-dependent time-delay of the maxima is measured to a precision of 30 ps. From that data, we find the cable propagation time of $(4.98 \pm 0.06)$ ns m$^{-1}$, a corresponding velocity factor of $(0.671 \pm 0.008)$, and a sensitivity to variations in the cable length of about 6 mm.

Discussion and Tweaks

The lock-in amplifier measures the difference signal $I_A$-$I_B$, that is, the difference between the signal created during cycle A and cycle B. This leads to the cancellation of the currents due to arbP, since the pump-pulse is the same in both parts of the cycle. Consequently, the interesting signal carrying the dynamics of the system due to the probe-pulse is enhanced.

The earliest implementation of the all-electronic pump-probe method used a mechanical chopper, effectively suppressing arbB [6]. In our implementation, this would be equivalent to setting arbB = 0 during cycle B [6]. A better signal is achieved, through the inverted voltage of arbB=-arbA, because we obtain data also for the second half of the entire cycle. However, since the preamplifier acts like a charge storage that is loaded and unloaded through the voltage induced current pulses, it may become overloaded through the voltage pulses of arbP. A remedy is the application of current-cancelling pump-pulses, either via an AC pulse (as shown above using the RF generator) or a positive/negative pulse [17]. Our waveform-sequencing method can be easily modified to overcome these issues that may arise if the ratio of the pump and probe-pulse amplitudes becomes too big, by simply changing the form of arbP.

The waveform-sequencing methods described here allows for a flexible design of arbitrarily shaped pump/probe-pulse trains. Our method will also work, if pulse-shaping is employed to compensate frequency dependent transmission losses to improve edge times [16], since our method is completely independent of the explicit shape of the arb.

Waveform-sequencing may also help overcome memory related limitations. For instance, a waveform containing the entire cycle A and B inside a single waveform as done previously can occupy a large portion of the waveform memory and it needs to be transferred to the AWG during the experiment. This may not only lead to an interruption of the AWG output but the permanent data transfer for different time-increments also creates noticeable overhead. Depending on the AWG memory, explicitly coding of an entire cycle A and B, may further require prolonging effective edge-times. No such limitations are present in our method. We achieve rising and falling edge times of 3 ns for any pulse-width at full sample-rate of 1 GSa/s. The shortest pulse-width is limited here only by the technical characteristics of our AWG.

We should also mention that any two voltage sources will have different voltage offsets that need to be compensated. This is particularly important, because as we switch between two nominally equivalent DC voltages of the STM control electronics and the AWG, we might effectively apply a different voltage. We therefore calibrate the voltage offsets by separately measuring two current-voltage traces at two impedances (for instance 1 GΩ and 100 MΩ). The crossing of the two I-V traces determines the zero-bias offset for every voltage source. As a verification, a correction of the offsets should lead to the same current or tip-height when switching from the STM control to the AWG.

We foresee the utility of our setup also for the anticipated advent of STM based pulsed-ESR methods. This is owed to the tremendous promise revealed in the study of individual electron and nuclear spins [12,18–20] through the introduction of ESR capable STM systems [15,21]. Although these STM based ESR studies relied on continuous wave excitation, the ongoing search for spin systems having long coherence times or the exploitation of clock-transitions [22–24] will ultimately lead to the introduction of pulsed ESR-STM methods that coherently control the quantum states of atomic and molecular qubits.

**Acknowledgements:** *Support from the Swiss National Science Foundation under project number PP00P2_176866 is appreciated. We thank Alexander Khajetoorians and Jan Gerritsen and for discussions and hospitality.*